\documentclass[aoas,MSNbibl,nameyear,dvips]{arximspdf}
\usepackage{dcolumn}
\usepackage{mathbh}
\usepackage{graphicx}

\doi{10.1214/12-AOAS602} 
\volume{7}
\issue{1}
\pubyear{2013}
\firstpage{588}
\lastpage{612}

\makeatletter

\newcommand{\iint}{\int\!\!\!\int}
\newcolumntype{d}[1]{D{.}{.}{#1}}

\newproclaim{definition}[theorem]{Definition}
\newproclaim{example}[theorem]{Example}
\def\cal{\mathcal}
\newcommand{\Cov}{\operatorname{Cov}}
\newcommand{\real}{\mathbb{R}}
\newcommand{\indicator}{\mathbh{1}}
\newcommand{\E}{\mathbb{E}}
\newcommand{\ba}{\mathbf{a}}
\newcommand{\bh}{\mathbf{h}}
\newcommand{\bs}{\mathbf{s}}
\newcommand{\bu}{\mathbf{u}}
\newcommand{\by}{\mathbf{y}}
\newcommand{\bv}{\mathbf{v}}
\newcommand{\bC}{\mathbf{C}}
\newcommand{\bb}{\mathbf{b}}
\newcommand{\bX}{\mathbf{X}}
\newcommand{\bbeta}{\bolds{\beta}}
\makeatother

\begin{document}
\begin{frontmatter}

\title{Daily minimum and maximum temperature simulation over complex
terrain\thanksref{T1}}
\thankstext{T1}{Supported by the NCAR Weather and Climate Assessment
Science Program and by NSF Coupled Natural Human Systems Program
Grant CNH-0709681 and NSF Earth System Models Program EaSM-1049099.
NCAR is sponsored by the National Science Foundation.}
\runtitle{Temperature simulation over complex terrain}

\begin{aug}
\author[A]{\fnms{William} \snm{Kleiber}\corref{}\ead[label=e1]{william.kleiber@colorado.edu}},
\author[B]{\fnms{Richard W.} \snm{Katz}}
\and
\author[C]{\fnms{Balaji} \snm{Rajagopalan}}
\runauthor{W. Kleiber, R.~W. Katz and B. Rajagopalan}
\affiliation{University of Colorado, National Center for Atmospheric
Research and University of Colorado}
\address[A]{W. Kleiber\\
Department of Applied Mathematics\\
University of Colorado\\
Boulder, Colorado\\
USA\\
\printead{e1}} 
\address[B]{R.~W. Katz\\
Institute for Mathematics Applied to Geosciences\\
National Center for Atmospheric Research\\
Boulder, Colorado\\
USA}
\address[C]{B. Rajagopalan\\
Department of Civil, Environmental\\
\quad and Architectural Engineering\\
University of Colorado\\
Boulder, Colorado\\
USA}
\end{aug}

\received{\smonth{6} \syear{2012}}
\revised{\smonth{8} \syear{2012}}

%
\begin{abstract}
Spatiotemporal simulation of minimum and maximum temperature is
a fundamental requirement for climate impact studies and hydrological or
agricultural models. Particularly over regions with variable orography,
these simulations are difficult to produce due to terrain driven
nonstationarity.
We develop a bivariate stochastic model for the spatiotemporal
field of minimum and maximum temperature.
The proposed framework splits the bivariate field into two
components of ``local climate'' and ``weather.'' The local climate
component is
a linear model with spatially varying process coefficients
capturing the annual cycle and yielding
local climate estimates at all locations, not only those within the observation
network. The weather component spatially correlates the bivariate simulations,
whose matrix-valued covariance function we estimate using a
nonparametric kernel smoother that retains nonnegative definiteness
and allows for substantial nonstationarity across the simulation domain.
The statistical model is augmented with a spatially varying nugget effect
to allow for locally varying small scale variability. Our model is applied
to a daily temperature data set covering the complex terrain of
Colorado, USA,
and successfully accommodates substantial temporally varying nonstationarity
in both the direct-covariance and cross-covariance functions.
\end{abstract}

%
\begin{keyword}
\kwd{Complex terrain}
\kwd{Gaussian process}
\kwd{nonstationary}
\kwd{minimum temperature}
\kwd{maximum temperature}
\kwd{multivariate covariance}
\kwd{simulation}
\kwd{stochastic weather generator}
\end{keyword}

\end{frontmatter}

\section{Introduction} \label{secintroduction}

Stochastic simulation of physical variables such as minimum or maximum
temperature, precipitation amount and solar radiation are often
required\vadjust{\goodbreak}
as inputs to physical models over varying types of topography.
Over plains regions, agricultural and crop models require daily minimum
and maximum temperature simulations at locations that typically do not
have direct observations. In mountainous regions, hydrological models
require stochastic weather realizations for runoff, snowmelt and watershed
modeling, as well as water resource planning and climate impact assessment
[\citet{kustas1994}, \citet{semenov1997}].

Stochastic weather generators (SWGs) are one approach to producing
simulations of daily weather; they are simply probability models whose
simulations are statistically similar to observations [\citet{wilks1999ppg}].
SWGs can loosely
be categorized into model-based [e.g., \citet{racsko1991}, \citet{richardson1981}]
and empirical approaches [e.g., \citet{lall1996}, \citet{rajagopalan1999}].
Often these
weather generators produce simulations only at locations with
observational data,
but modern physical models require gridded daily weather. Hence, recent
research has been directed toward generating spatially consistent SWGs
that are available at and between observation locations [\citet
{wilks1999}, \citet{kleiber2012}]. Herein we focus on a model-based approach to minimum
and maximum temperature simulation over a mix of complex terrain and
relatively homogeneous terrain simultaneously.

Spatially consistent simulation over most agricultural regions can be
accommodated using isotropic or stationary models that are appropriate
for regions with relatively constant or slowly changing topography.
Domains with highly variable terrain, in particular, mountainous domains,
are challenging for the majority of univariate spatial models due to
substantial nonstationarity of physical processes in these areas.
Weather over complex terrain is highly variable due to topography;
for example, at high elevations in the northern hemisphere,
north facing slopes tend to be cooler
than lower elevations and south facing slopes and valleys can create
their own
micro-climate relative to the surrounding high elevation.
These conspire to produce intricate spatial
variability that is hard for models to capture. A typical approach is to
partition the space into homogeneous regions and model each region separately.
While a number of statistical nonstationary spatial models have been
proposed for
univariate fields [\citet{fuentes2002},
\citet{haas1990},
\citet{higdon1998},
\citet{kim2005},
\citet{paciorek2006},
\citet{pintore2006},
\citet{sampson1992},
\citet{stroud2001}],
fewer are available for multivariate spatial simulation,
which is of key concern for simultaneous minimum and maximum
temperature simulation
[\citet{gelfand2004},
\citet{jun2011},
\citet{kleiber2012nsmm},
\citet{shaddick2002}].

%
\begin{figure}

\includegraphics{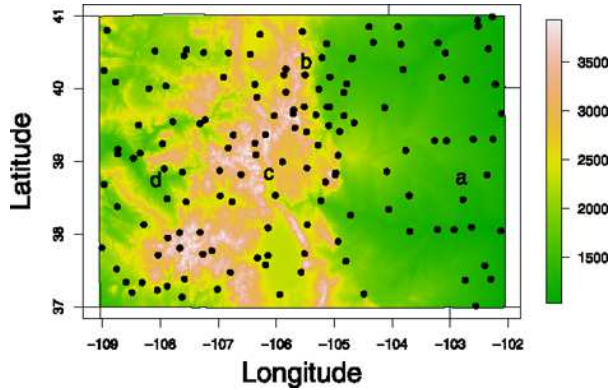}

\caption{Map of elevations in Colorado (in meters) and the 145
locations used
from the Global Historical Climatology Network. Four locations we later use
for cross-validation are denoted \textup{(a)}~Kit Carson, \textup
{(b)} Estes Park,
\textup{(c)} Buena Vista and \textup{(d)} Delta.} \label{fignetwork}
\end{figure}


Some literature in geography and the atmospheric sciences is
concerned with deterministic interpolation of observed
weather variables, often over domains with complex terrain
[\citet{daly1994},
\citet{hijmans2005},
\citet{hutchinson1995},
\citet{legates1990},
\citet{price2000},
\citet{running1987},
\citet{thornton1997},
\citet{willmott1995}].
The common theme among these approaches
is the inclusion of high resolution digital elevation maps as well
as other physical information such as slope and aspect to deterministically
interpolate meteorological variables.
While most of these models are sophisticated physical interpolation schemes,
they [apart from \citet{thornton1997}] are chiefly concerned with
monthly or annual average
quantities, and do not produce stochastic realizations of daily weather,
which is our primary interest. These schemes are also typically ad hoc,
and are not based on a formal statistical model.

Stochastic interpolation and simulation of physical variables has
persistent interest in the statistics literature. Often, precipitation
holds the primary interest, as its mixed discrete-continuous and skewed
nature pose substantial challenges [\citet{ailliot2009},
\citet{allcroft2003},
\citet{brown2001},
\citet{durban2001},
\citet{hughes1999},
\citet{sanso2000}].
However, recent authors
have acknowledged the difficulties of temperature modeling in complex
terrain [\citet{paciorek2006}], and \citet{gelfand2005} is one of few
to simultaneously model temperature and precipitation.

The study domain in this paper is the state of Colorado.
Figure~\ref{fignetwork} illustrates the challenging
terrain of Colorado, with eastern plains dipping to a minimum elevation
of approximately 1000~m and the Rocky Mountains of central Colorado peaking
out at above 4000~m. The front range, the ridge separating the Rocky
Mountains from the eastern plains (running north-south on approximately
the $-105^{\circ}$ longitude line), is especially difficult to
accommodate using
the currently available multivariate covariance models, most of which
are isotropic models, and do not allow for sudden boundaries or even
gradually evolving spatial structures across a domain.
The 145 locations shown in Figure~\ref{fignetwork} are
a subset of stations from the Global Historical Climatology
Network Database [GHCND; \citet{peterson1997}].
Daily observations of minimum and maximum temperatures are available
between a time period of at most 1893 through 2011. Associated with each
observation is a quality flag provided
by the GHCND; we removed all flagged observations to avoid poor quality
observations.

In this paper we propose a framework for bivariate stochastic temperature
simulation that splits the model into two components. The first component
represents local climate, allowing the average behavior of minimum and
maximum temperature to vary with location, which is of critical concern
in regions such as Colorado with the average behavior of temperature
in the Rocky Mountains being vastly different than that over the
eastern plains.
The second component can be interpreted as
daily weather, yielding local variability in space and time, and
preserving the
spatial correlation between both processes.

\section{Stochastic model} \label{secmodel}

Consider the bivariate process of minimum temperature, $Z_N(\bs,t)$,
and maximum temperature, $Z_X(\bs,t)$, at location $\bs\in\real^2$
on day $t=1,\ldots,T$. Our model for the bivariate process is
%
\begin{eqnarray}
Z_N(\bs,t) &=& \bbeta_N(\bs)'
\bX_N(\bs,t) + W_N(\bs,t), \label{eqNmodel}
\\
Z_X(\bs,t) &=& \bbeta_X(\bs)'
\bX_X(\bs,t) + W_X(\bs,t) \label{eqXmodel}.
\end{eqnarray}
The vector of coefficients $\bbeta_i(\bs) = (\beta_{0i}(\bs),\beta
_{1i}(\bs),
\ldots,\beta_{pi}(\bs))'$, for $i=N,X$, may be of different length for
minimum and maximum temperatures, allowing for distinct sets of
covariates, although
for notational simplicity we assume both processes share the same
number of
covariates, $p+1$. The covariates $\bX_i(\bs,t) = (X_{0i}(\bs
,t),\ldots,X_{pi}(\bs,t))'$
typically involve autoregressive and seasonality terms and, if available,
can contain additional information such as regional climate model output.
It is convenient to view the models of (\ref{eqNmodel}) and (\ref{eqXmodel})
as a sum of ``local climate'' plus ``weather.'' The local climate is
dependent on spatially and temporally varying covariates, and whose coefficients
vary across the domain, allowing for the relative influence of each
covariate to
depend on location. The weather terms, $W_N(\bs,t)$ and $W_X(\bs,t)$,
capture small
scale variability and correlate the bivariate temperature process across
space.

\subsection{Local climate component} \label{secclimate}

The coefficients $\beta_{ki}(\bs)$, for $i=N,X$ and $k=0,\ldots,p$,
allow the average
behavior of temperature to vary with location. This is crucially important
in areas of complex terrain or over large domains where variable orography
and general circulation patterns give rise to varying climate [\citet{chandler2005},
\citet{johnson2000},
\citet{kleiber2012}]. \citet{pepin2002} point out
that climate change
trends in Colorado are highly dependent on the terrain.
Direct estimates of these coefficients are usually only available at locations
within the observation network, so we model the coefficients as spatial Gaussian
processes. In particular, we suppose $\beta_{ki}(\bs)$ has mean $\mu_{ki}$
and Mat\'ern covariance augmented with a nugget effect,
with variance parameter $\sigma_{ki}^2$,
range $a_{ki}$, smoothness $\nu_{ki}$ and nugget effect $\tau_{ki}^2$
[\citet{guttorp2006}]. The goal of a spatial model for the coefficients
$\beta_{ki}(\bs)$ is for interpolation from the observational network
locations
to a chosen grid. The Mat\'ern is an isotropic covariance function that
is especially useful for kriging [\citet{stein1999}]. One might
consider using a nonstationary function for the coefficient covariance
model, but in our experience (see the example section below), the
simpler stationary model works well for local climate interpolation.

For Colorado, we use the following covariates:
%
\begin{eqnarray}
\label{eqcovariates}\qquad \bX_N(\bs,t) = \biggl(1,\cos\biggl(
\frac{2\pi t}{365} \biggr), \sin\biggl(\frac{2\pi t}{365} \biggr),Z_X(
\bs,t-1), Z_N(\bs,t-1),r_t \biggr)',
\end{eqnarray}
with the corresponding case for $\bX_X(\bs,t)$ reversing indices $N$
and $X$.
The harmonics allow for seasonality in minimum and maximum
temperatures, and we include bivariate autoregressive terms to account for
temporal persistence of temperature. The final
covariate, $r_t$, is a linear drift of length $T$ between $-1$ and 1
(for numerical stability), which we include to control for temperature
trends over the 119 year period of our data set, noting that these trends
do not necessarily reflect global warming.
These covariates were selected using a BIC criterion at all individual stations;
that is, fitting a model to each location independently,
the model with all of the above covariates had the smallest BIC
value for all stations within the GHCND in Colorado, as compared to
any subset of the selected covariates. We considered
models with higher order harmonics and autoregressive lags,
but the results were nearly identical to
those presented below, hence, we favor the simpler set of covariates.

Suppose we observe the bivariate process $(Z_N(\bs,t),Z_X(\bs,t))'$
at locations
$\bs=\bs_1,\ldots,\bs_n$ and time points $t = 1,\ldots,{T}$. At
each location within the observation network, we estimate local parameters
$\hat{\beta}_{ki}(\bs)$ by ordinary least squares. These estimates
have low
uncertainty; in the Colorado network, the location with the sparsest
observational record
still has more than 10,000 available observations. Conditional on the estimates
$\hat{\beta}_{ki}(\bs)$, we estimate the spatial Gaussian process parameters
$\mu_{ki}, \sigma_{ki}^2, a_{ki}, \nu_{ki}$ and $\tau_{ki}^2$ by maximum
likelihood, exploiting the Gaussian process assumption. These spatially
varying coefficients models [\citet{gelfand2003}] have been used
for probabilistic forecasting, with a similar two-step estimation procedure
[\citet{kleiber2011mwr},
\citet{kleiber2011jasa}].

At an arbitrary location $\bs_0$, not necessarily within the
observation network,
we spatially interpolate the estimates $\hat{\bbeta}_{ki} = (\hat
{\beta}_{ki}(\bs_1),
\ldots,\hat{\beta}_{ki}(\bs_n))'$ via\vadjust{\goodbreak}
kriging [\citet{cressie1993}]. In particular, the kriging estimator is
\[
\hat{\beta}_{ki}(\mathbf{s}_0) = \mathbf{c}' \Sigma^{-1} (\hat{\bbeta}_{ki} - \mu_{ki} \mathbf{1}) +
\mu_{ki}
\]
and the interpolation variance is
\[
\sigma_{ki}^2 + \tau_{ki}^2 -
\mathbf{c}' \Sigma^{-1} \mathbf{c},
\]
where $\mathbf{1}$ is a vector of 1s of length $n$, $\mathbf{c}' = (\Cov(\beta_{ki}(\bs
_0),\beta_{ki}(\bs_1)),\ldots,\break
\Cov(\beta_{ki}(\bs_0),\beta_{ki}(\bs_n)))$ and $(\Sigma)_{j,\ell
} =
\Cov(\beta_{ki}(\bs_j),\beta_{ki}(\bs_\ell))$ for $j,\ell
=1,\ldots,n$.
As kriging is an exact interpolator, when $\bs_0 = \bs_\ell$ for any
$\ell=1,\ldots,n$, the interpolator returns the ordinary least squares
estimate $\hat{\beta}_{ki}(\bs_\ell)$.

In the next section we exploit a nonparametric estimator of the covariance
function for the bivariate weather process. Key to the nonparametric estimator
being consistent is a large number of realizations of the process
[\citet{kleiber2012nsmm}] which
we have available for the residual weather processes, whereas the
coefficient processes
of the local climate component have only one realization. Hence, we
favor the
parametric model with a two-step estimation procedure for local climate.

\subsection{Weather component} \label{secweather}

To simulate spatially correlated fields of minimum and maximum temperatures
consistent with observed spatial patterns, we require a bivariate
spatial model
for $W_N(\bs,t)$ and $W_X(\bs,t)$. In particular, we model these weather
processes as a zero-mean bivariate spatial Gaussian process indexed by
day of the year.
For locations $\mathbf{x},\by$, and arbitrary time
point $t$, the bivariate covariance model is
%
\begin{eqnarray}
&\Cov\bigl(W_i(\mathbf{x},t),W_j(\by,t+1)\bigr)
= 0,
\label{eqWtime}
\\
&\Cov\bigl(W_i(\mathbf{x},t),W_i(\by,t)\bigr) =
C_{ii}\bigl(\mathbf{x},\by,d(t)\bigr) + \tau
_i(\mathbf{x},
\by)^2 \indicator_{[\mathbf{x} = \mathbf{y}]}, \label{eqWcov}
\\
&\Cov\bigl(W_i(\mathbf{x},t),W_j(\by,t)\bigr) =
C_{ij}\bigl(\mathbf{x},\by,d(t)\bigr)\qquad \mbox{for } i\not=j\label{eqWcross}
\end{eqnarray}
for $i,j=N,X$, where $d(t) \in\{1,\ldots,365\}$ is just the calendar
day of the year on which time point $t$ falls.


The covariance model of (\ref{eqWtime}), (\ref{eqWcov}) and (\ref{eqWcross})
implies some important assumptions. First,
we assume temporal dependence has been accounted for in the
local mean function (e.g., via autoregressive terms) so that the weather
process is temporally independent, hence (\ref{eqWtime}).
Indeed, exploratory plots such as autocorrelation functions and
empirical covariance functions indicate the bivariate autoregression of
(\ref{eqcovariates}) is sufficient to account for the temporal persistence
of temperature in Colorado; see the example section below.
Second, the covariance and cross-covariance functions $C_{ii}(\mathbf{x},\by,d(t))$ and
$C_{ij}(\mathbf{x},\by,d(t))$ depend on the day of
year, allowing the bivariate
process to have seasonally dependent second-order structure.

In (\ref{eqWcov}), $\tau_i(\mathbf{x})^2 = \tau
_i(\mathbf{x},\mathbf{x})^2$ is
a local nugget effect,
accounting for small scale variability as well as measurement
error. In the geostatistical literature, $C_{ii}(\mathbf{x},\mathbf{x},d(t))$ is often termed\vadjust{\goodbreak}
the marginal variance, while $C_{ii}(\mathbf{x},\mathbf{x},d(t)) +
\tau_i(\mathbf{x})^2$ is called
the sill, that is, the total variance at a given location [\citet{cressie1993}].
Unlike most geostatistical models [\citet{christensen2011} being a notable
departure],
we allow the nugget effect to vary with
location, as we expect the small scale variability to be highly
dependent on
orography.

At any fixed time point $t$ [i.e., calendar day $d(t)$],
we require the matrix-valued covariance function
%
\begin{equation}
\bC\bigl(\mathbf{x},\by,d(t)\bigr) = \pmatrix{
C_{NN}
\bigl(\mathbf{x},\by,d(t)\bigr) & C_{NX}\bigl
(\mathbf{x},\by,d(t)\bigr)
\vspace*{2pt}\cr
C_{XN}\bigl(\mathbf{x},\by,d(t)\bigr) &
C_{XX}\bigl(\mathbf{x},\by,d(t)
\bigr)}
\label{eqmatrixcov}
\end{equation}
to be a nonnegative definite matrix function. Specifically, at arbitrary
locations $\bs_1,\ldots,\bs_n$, the covariance matrix of the random vector
\[
\bigl(W_N(\bs_1,t),W_X(
\bs_1,t),W_N(\bs_2,t),W_X(
\bs_2,t),\ldots,W_N(\bs_n,t),W_X(
\bs_n,t)\bigr)',
\]
which is made up of blocks $\bC(\bs_k,\bs_\ell,d(t))$, must be
nonnegative definite.

Over regions with complex terrain, temperature observations can exhibit
substantial nonstationarity [\citet{paciorek2006}]. While some multivariate
spatial models that can account for nonstationarity are available
[e.g., \citet{gelfand2004},
\citet{kleiber2012nsmm}], these are parametric
models with locally varying
parameter functions that are difficult to estimate. We aim to exploit
the large number of replications and reasonably well covered
observation network of the
GHCND over Colorado, and propose a nonparametric estimator of the matrix-valued
covariance function that retains nonnegative definiteness.
In particular, suppose the bivariate process is observed
at locations $\bs_k, k=1,\ldots,n$, and times $t=1,\ldots,T$.
Then our nonparametric estimator of $C_{ij}(\mathbf{x},\by,d(t_0))$ in (\ref{eqWcov})
and (\ref{eqWcross}),
at location pair $(\mathbf{x},\by)$ and time point
$t_0$ is
%
\begin{eqnarray}
\label{eqnp} &&\hat{C}_{ij}\bigl(\mathbf{x},\by
,d(t_0)\bigr)
\nonumber\\
&&\qquad  =\Biggl(\sum_{t=1}^{T} \sum_{k=1}^n \sum_{\ell=1}^n
K_{\lambda_t} \bigl(\bigl\|d(t_0),d(t)\bigr\|_d \bigr) K_\lambda\bigl(\|\mathbf{x}-\bs_k\| \bigr)
K_\lambda\bigl(\|\by-\bs_\ell\| \bigr)
\nonumber
\\[-8pt]
\\[-8pt]
\nonumber
&&\hspace*{169pt}\qquad\quad{}\times W_i(\bs_k,t) W_j(\bs_\ell,t)\Biggr)\\
&&\qquad\quad{}\bigg/\Biggl(\sum_{t=1}^{T} \sum_{k=1}^n \sum
_{\ell=1}^n
K_{\lambda_t}\bigl(\bigl\|d(t_0),d(t)\bigr\|_d\bigr) K_\lambda\bigl(\|\mathbf{x}-\bs_k\|\bigr)
K_\lambda\bigl(\|\by-\bs_\ell\|\bigr)\Biggr)\nonumber
\end{eqnarray}
for $i,j=N,X$. Here, $K_\lambda$ is a kernel function with bandwidth
$\lambda$, and
we use $K_\lambda(\|\bh\|) = (1/\lambda)\exp(-\|\bh\|/\lambda)$.
We use the Euclidean norm
$\|\cdot\|$, and the distance function
$\|\cdot,\cdot\|_d$ is the distance between days of the year so that
$\| d_1,d_2\|_d = |d_1-d_2|$ for $|d_1-d_2|\leq182$ and
$\| d_1,d_2\|_d = |365-|d_1-d_2||$ for $|d_1-d_2|>182$,
where $d_1,d_2=1,\ldots,365$,\vadjust{\goodbreak} for example, $\| 1, 365\|_d = 1$.
Occasionally $Z_i(\bs_k,t)$ [and subsequently $W_i(\bs_k,t)$]
is not available in practice due to instrument
failure or disruptions in communications. The estimator we use
operationally is a slightly modified version of (\ref{eqnp}),
where we make the convention $W_i(\bs,t)
\indicator_{[W_i(\mathbf{s},t)\ \mathrm{is\ observed}]} =
0$ when $W_i(\bs,t)$ is missing. It is convenient to define
the single-time-point smoothed empirical covariance
function\looseness=-1
%
\begin{eqnarray}
\label{eqnpR}
&&\hat{R}_{ij}(\mathbf{x},\by,t)
\nonumber\\[-3pt]
&&\qquad= \Biggl(\sum_{k=1}^n \sum_{\ell=1}^n
K_\lambda\bigl(\|\mathbf{x}-\bs_k\| \bigr)K_\lambda\bigl(\|\by
-\bs_\ell\| \bigr)
W_i(\bs_k,t) W_j(\bs_\ell,t)\nonumber\\[-3pt]
&&\hspace*{94pt}{}\times
\indicator_{[W_i(\mathbf{s}_k,t)\ \mathrm{is\ observed}]}
\indicator_{[W_j(\mathbf{s}_\ell,t)\ \mathrm{is\ observed}]}\Biggr)\\[-3pt]
&&\qquad\quad{}\bigg/ \Biggl(
\sum_{k=1}^n \sum_{\ell=1}^n
K_\lambda\bigl(\|\mathbf{x}-\bs_k\|\bigr) K_\lambda\bigl(\|\by
-\bs_\ell\|\bigr)\nonumber\\[-3pt]
&&\hspace*{50pt}\qquad\quad{}\times
\indicator_{[W_i(\mathbf{s}_k,t)\ \mathrm{is\ observed}]}
\indicator_{[W_j(\mathbf{s}_\ell,t)\ \mathrm{is\
observed}]}\Biggr).\nonumber
\end{eqnarray}\looseness=0
Notice $\hat{R}_{ij}(\mathbf{x},\by,t)$ is just a
(spatially) smoothed empirical
covariance function over the available observations on day $t$.
$\hat{R}_{ij}(\mathbf{x},\by,t)$ is a nonnegative
definite multivariate covariance
function, a property we show in the \hyperref[app]{Appendix}.
Our adjusted version of (\ref{eqnp}) that accounts for missing observations
then is
%
\begin{equation}
\label{eqnpmissing} \hat{C}_{ij}\bigl(\mathbf{x},\by,d(t_0)
\bigr) = \frac{\sum_{t=1}^T K_{\lambda_t} (\|d(t_0),d(t)\|_d )
\hat{R}_{ij}(\mathbf{x},\by,t)} {
\sum_{t=1}^{T} K_{\lambda_t}(\|d(t_0),d(t)\|_d)}.
\end{equation}
As $\hat{C}_{ij}(\mathbf{x},\by,d(t_0))$ is a
positively weighted
linear combination of multivariate
covariance functions, it is again nonnegative definite. The
estimator for missing observations (\ref{eqnpmissing}) reduces
to the original estimator (\ref{eqnp}) when no observations are missing
and, hence, (\ref{eqnp}) is also nonnegative definite.

The estimator (\ref{eqnp}) is a smoothed version of daily empirical covariance
matrices. The first level of smoothing yields an estimate of spatial covariance
at any arbitrary location pairs in the domain. The temporal smoothing
shares information
between adjacent time points, where we assume that spatial covariance
on a given day is similar to that in a short period leading up to that day,
and in a short period following that day.
This estimator is a generalization of kernel smoothed empirical covariance
estimators considered by \citet{oehlert1993}, \citet{guillot2001} and
\citet{jun2011mwr}
to the multivariate process setting evolving across time.

We estimate the time bandwidth $\lambda_t$ by predictive leave-one-out
cross-validation, leaving out local empirical variance estimates.
The estimated bandwidth for time is $\hat{\lambda}_t =
7.8$ days. We use cross-validation for the temporal bandwidth, as we
assume the temporal evolution of spatial covariance is\vadjust{\goodbreak} slowly evolving
across time, for example,~we do not expect a sharp change in spatial
covariance between
June~1 and June~2. In our experience, using cross-validation for the spatial
bandwidth parameter $\lambda$ oversmooths the spatial covariance function.
When kernel smoothing a mean function, cross-validation is generally
acknowledged
to yield more variability than is expected for a smoothly varying mean function,
and typically the bandwidth must be inflated [\citet{wand1995}].
However, this experience is under the assumption that the mean function
is varying smoothly across the domain, and in regions of complex
terrain we
expect the opposite behavior, where sharp boundaries of the covariance
function may exist due to sudden changes in elevation. For example,
cross-validation implies the optimal spatial bandwidth is 75~km, which
implies an effective range of the kernel function (i.e., up to $5\%$ weight)
of approximately 225 km, greatly oversmoothing regions such as the
San Luis Valley in southern
Colorado, at approximately 100 km across. Hence, we choose a bandwidth
such that the effective distance of the kernel
function coincides with the $5\%$ quantile of all intersite distances
(62 km); the heuristic argument is that, for approximately evenly
distributed observation locations, the covariance estimator at a given
location uses the nearest $5\%$ of available network locations
and down-weights remote locations;
this ad hoc criterion implies a spatial bandwidth of $\hat{\lambda} =
22$~km.

The estimator (\ref{eqnp}) is asymptotically unbiased for
$C_{ij}(\mathbf{x},\by,d(t_0))$
when the domain sample size increases and
the bandwidth decreases to zero sufficiently quickly. A short
argument is given in the \hyperref[app]{Appendix}.
In fact, it can be shown that the estimator is consistent for
$C_{ij}(\mathbf{x},\by,d(t_0))$,
using arguments similar to those of \citet{kleiber2012nsmm}, but this is
beyond the scope of the present paper.

All that remains to be estimated is the local nugget effect $\tau
_i(\bs)^2$.
At each observation location $\bs_k, k=1,\ldots,n$, and time point
$t=1,\ldots,T$, let $W_i(\bs_k,t)$ be the estimated residual $Z_i(\bs
_k,t) -
\hat{\bbeta}_i(\bs_k)'\bX_i(\bs_k,t)$. Define the local empirical variance
on day $d=1,\ldots,365$ as
\[
\hat{\sigma}_{i}(\bs_k,d)^2 =
\frac{1}{\# \{ t | d(t) = d \}} \sum_{\{ t | d(t) = d \}} W_i(
\bs_k,t)^2,
\]
where $\#$ denotes cardinality of the set, with the natural redefinition
for missing values of $W_i(\bs_k,t)$. Intuitively, a good
estimator for $\tau_i(\bs_k)^2$ is
%
\begin{equation}
\label{eqtau} \hat{\tau}_i(\bs_k)^2 =
\frac{1}{365} \sum_{d=1}^{365} \bigl(\hat{
\sigma}_i(\bs_k,d)^2 - \hat{C}_{ii}(
\bs_k,\bs_k,d) \bigr),
\end{equation}
since, by the law of large numbers, $\hat{\sigma}_{i}(\bs_k,d)^2
\rightarrow C_{ii}(\bs_k,\bs_k,d) + \tau_i(\bs_k)^2$, where the convergence
is taken as $T\rightarrow\infty$, and by the argument in the \hyperref[app]{Appendix},
$\hat{C}_{ii}(\bs_k,\bs_k,d) \rightarrow C_{ii}(\bs_k,\bs_k,d)$.
While theoretically appealing, in practice, due to the smoothing in
$\hat{C}_{ii}$, at some locations the estimate $\hat{\tau}_i(\bs
_k)$ is
negative. Hence, in similar spirit we use (\ref{eqtau}),
but set the invalid estimates to zero.\vadjust{\goodbreak}

Estimates of $\tau_{i}(\bs)^2$ are gathered at arbitrary locations,
that is, not
necessarily within the observation network, by imposing a probabilistic
spatial structure on $\tau_i(\bs)$. In particular,
we model $\tau_i(\bs)$ as a Gaussian process with spatially constant mean
and Mat\'ern covariance function, augmented with a nugget effect. Just
as for the spatial parameters of the $\beta_{ki}(\bs)$, we estimate
the spatial parameters of $\tau_i(\bs)$ by maximum likelihood, conditional
on the estimates $\{\hat{\tau}_i(\bs_k)\}_{k=1}^n$. While the estimates
$\hat{\tau}_i(\bs_k)$ at observation locations are always valid, the
kriging interpolator of $\tau_i(\bs)$ may occasionally take on very
small negative
values; in our example below we did not experience such an issue, but
in other domains these degenerate estimates may be artificially set to zero.

\section{Minimum and maximum temperature in Colorado} \label{secexample}

We fit our model to the data from the 145 GHCND locations shown in
Figure~\ref{fignetwork}. For simplicity, we removed all leap days from the
119 years of available data, so that
each year has 365 days. Using all available data, we fit local climate
parameters by ordinary least squares and estimate temporally varying
multivariate spatial covariances using the nonparametric estimator
(\ref{eqnpmissing}) applied to the observed residuals. We then simulate
the bivariate process for a 119 year trajectory to compare to the observed
bivariate series. The first day's (January 1, 1893) simulation requires
autoregressive
terms in~(\ref{eqNmodel}) and~(\ref{eqXmodel}); we initialize using
the climatological domain average of minimum and maximum temperatures
on December~31.
The resulting simulations are masked to share the same missing value
pattern as the observations.

%
\begin{figure}

\includegraphics{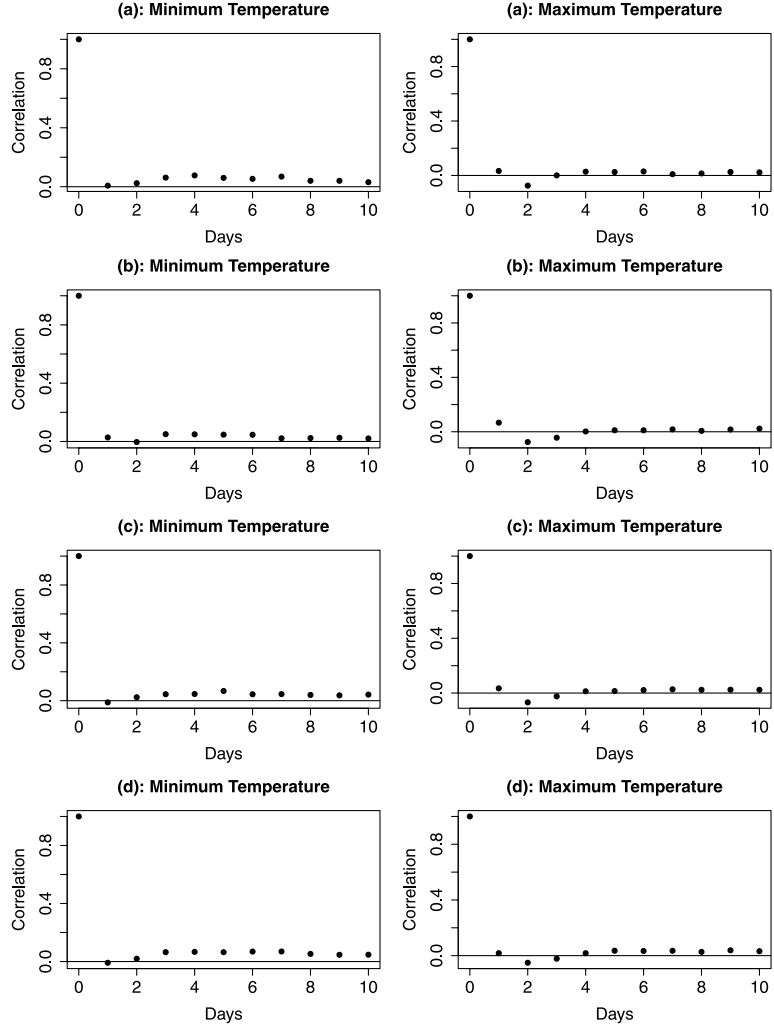}

\caption{Empirical autocorrelation functions for minimum and maximum
temperature residuals at locations
\textup{(a)} Kit Carson, \textup{(b)} Estes Park, \textup{(c)} Buena
Vista and \textup{(d)} Delta.}\label{figacfs}
\end{figure}


Recall the assumption implied by equation (\ref{eqWtime}), where
we assume temporal dependence has been accounted for in the
local mean function via the bivariate autoregression.
Figure~\ref{figacfs} contains empirical autocorrelation functions
for the observed residuals $W_N(\bs,t)$ and $W_X(\bs,t)$ at four
network stations, shown in Figure~\ref{fignetwork}.
These locations we view as representative of four
distinct regimes of Colorado: eastern plains (a, Kit Carson),
front range (b, Estes Park), Rocky Mountains (c, Buena Vista) and
the western slopes (d, Delta). It is evident that the bivariate
autoregression accounts for the majority of temporal persistence
in temperature; the maximal lag-1 autocorrelation coefficient for
the residual processes at these four stations is $0.06$ at Estes Park, whereas
all other coefficients are less than or equal to $0.03$.

%
\begin{figure}

\includegraphics{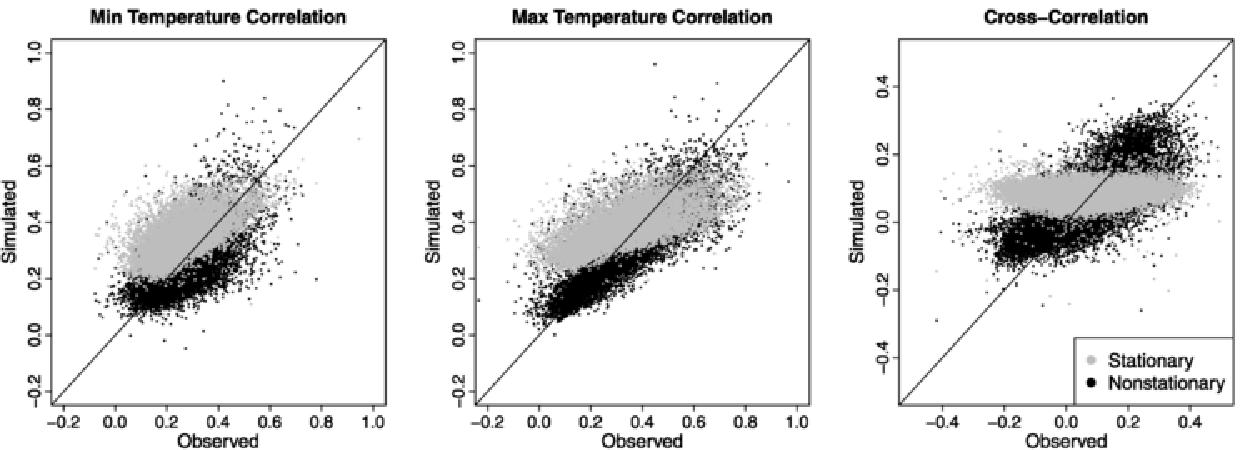}

\caption{Scatterplots comparing empirical pairwise station correlation
to simulated correlations using a bivariate stationary model (grey dots)
or the nonstationary nonparametric model (black dots)
over the summer months (JJA). The diagonal line indicates perfect agreement
between model and empirical correlations.}\label{figpairwise}
\end{figure}


To motivate the flexibility of the nonparametric estimator (\ref{eqnp}),
we compare it to
a state-of-the-art isotropic bivariate spatial model. In particular, we
fit a bivariate Mat\'ern model [\citet{gneiting2010},
\citet{apanasovich2012}]
augmented with a nugget effect, where
%
\begin{eqnarray}\quad
C_{ii}(\mathbf{x},\by,t) &=& \frac{1}{2^{\nu
_i-1}\Gamma(\nu_i)} \bigl(a_i\|\mathbf{x}-\by
\|\bigr)^{\nu_i} \mathrm{ K}_{\nu_i}\bigl(a_i\|\mathbf{x}-\by\|
\bigr) +
\tau_i^2\indicator_{[\mathbf{x}=\mathbf{y}]} ,
\\
\quad\qquad C_{NX}(\mathbf{x},\by,t) &= &\rho_{NX} \frac
{1}{2^{\nu_{NX}-1}\Gamma(\nu_{NX})}
\bigl(a_{NX}\|\mathbf{x}-\by\|\bigr)^{\nu_i} \mathrm{ K}_{\nu
_{NX}}\bigl(a_{NX}
\|\mathbf{x}-\by\|\bigr)\vadjust{\goodbreak}
\end{eqnarray}
for $i=N,X$, where $\mathrm{ K}_\nu$ is a modified Bessel function of the
second kind of order $\nu$,
and $\nu_{NX} = (\nu_N+\nu_X)/2$ and $a_{NX} = \min(a_N,a_X)$.
We fit the parameters by maximum likelihood, viewing each bivariate
estimated residual $(W_N(\bs,t),W_X(\bs,t))$ as independent across time.
In the stochastic weather simulation literature, it is customary to fit
separate models for each season. While our nonparametric estimator
is available on any day, to facilitate comparisons to the bivariate
Mat\'ern, we fit both models
to only the summer months (JJA), and compare
empirical to simulated correlations and cross-correlations under both
the isotropic and nonparametric models; Figure~\ref{figpairwise} displays
these results. The stationary model tends to overestimate spatial
correlation for both minimum and maximum temperatures, whereas our
nonstationary model adequately captures low and high correlations
simultaneously. The third panel of Figure~\ref{figpairwise} shows
empirical against simulated cross-correlations. Substantial nonstationarity
of cross-correlation across Colorado is well modeled by our
nonparametric approach, but the stationary model clearly fails,
putting most cross-correlations at around $0.10$, whereas the
empirical estimates suggest the true cross-correlations should vary
between $-0.10$ and $0.40$.

%
\begin{figure}

\includegraphics{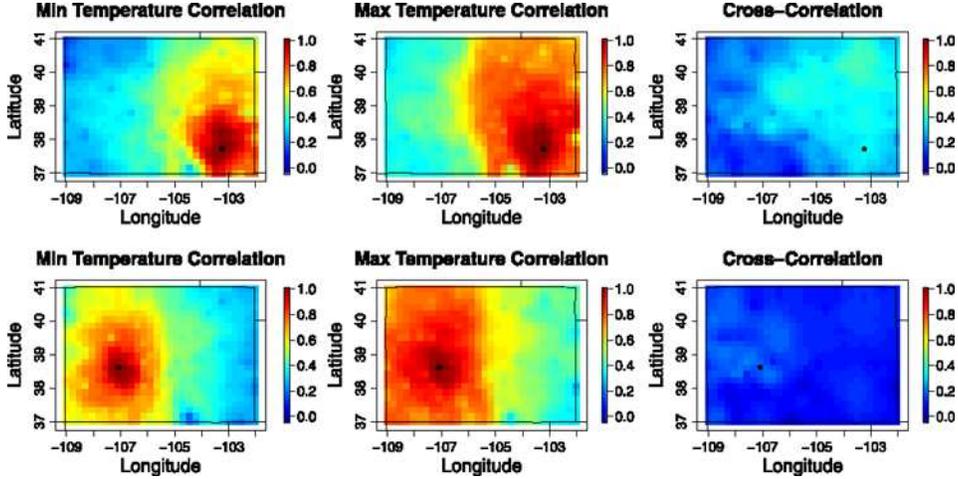}

\caption{Plots of spatial correlation and cross-correlation functions,
$C_{ij}(\bs_0,\cdot,d(t))$, on $d(t)=$ June~1 where $\bs_0$ is a
grid location in the eastern plains (top row) or a grid location in the
Rocky Mountains (bottom row), with grid locations indicated by black dots.
Each pixel's color indicates the model estimated spatial correlation
between the pixel location and the dot.}
\label{figspatialcor}\vspace*{-3pt}
\end{figure}


Our nonparametric matrix covariance estimator (\ref{eqnp}) accommodates
nonstationary behavior of the multivariate process.
Figure~\ref{figspatialcor}
shows two covariance functions on June 1, one whose first argument is based
at a grid location in the eastern plains of Colorado, and the second
covariance function whose first argument is based at a grid location in
the Rocky Mountains. The top row is the covariance function for
the plains-based grid location; particularly for maximum temperature, and
lesser so for minimum temperature, there is strong positive
within variable correlation throughout the plains region, suggesting that
maximum temperatures are highly correlated across the plains. At
the front range boundary (approximately $-105^\circ$ longitude),
there is a sharp drop off in spatial correlation from approximately
$0.80$ over the plains to $0.40$ in the Rocky Mountains.
This is due to the fact
that temperature is more highly correlated within the two main types
of topography of Colorado, either the plains or mountains, but not
between the two types. Hence, our estimator is
able to capture the sharp boundary between the eastern plains and
Rocky Mountains for within variable spatial correlation.
Our estimator also identifies the positive cross-correlation
between minimum and maximum temperatures in the plains, but allows
the two processes to be effectively independent over the Rocky
Mountains.\vadjust{\goodbreak}
This nonstationarity of cross-correlation is very difficult
to accommodate using extant models, and has only been recently
acknowledged in the literature [\citet{kleiber2012nsccc}].

%
\begin{figure}[b]\vspace*{-3pt}

\includegraphics{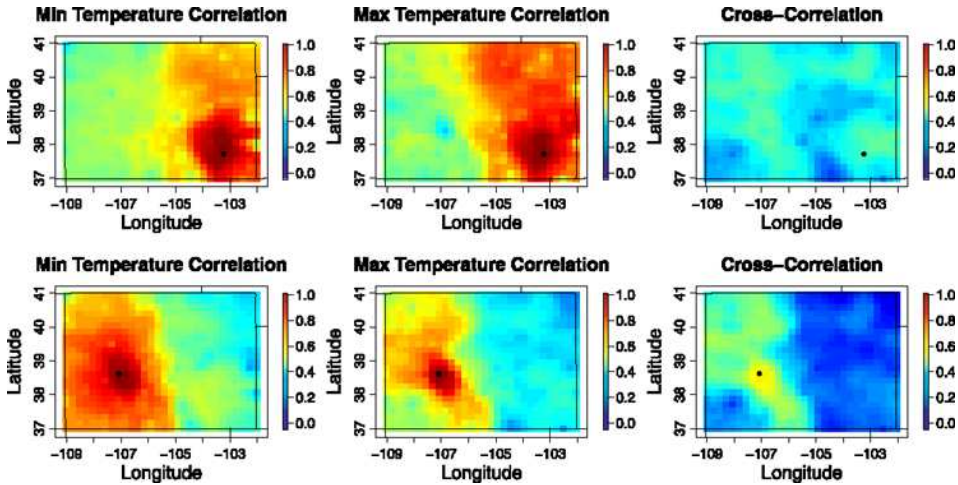}

\caption{Same as Figure \protect\ref{figspatialcor}, except for
$d(t)=$ January 1 instead of June 1.} \label{figspatialcorjan}
\end{figure}


Not only does our estimator allow for substantial nonstationarity,
the amount and type of nonstationarity is allowed to vary across
time. Figure~\ref{figspatialcorjan} shows the same plots\vadjust{\goodbreak} of
spatial direct and cross-correlation on January 1, during winter,
as opposed to the summer estimates of Figure~\ref{figspatialcor}.
In terms of direct covariance, we see the length scale of minimum
temperature correlation drastically increase for both the plains-
and mountain-based grid locations. 
In the plains, the spatial correlation structure of
maximum temperature is similar during both the winter and summer;
on the other hand, this spatial correlation in the mountainous
region over winter has a substantially different pattern than
over summer. The correlation structure of the weather
component for maximum temperature in the Rocky Mountains is
clearly nonstationary, implying lower correlation between the
example grid point and the southwestern slopes of the Rockies,
but having higher correlation along a northwest to southeast
transect along the western slopes and through the Rocky Mountains;
this pattern makes sense climatologically, as the band of
high correlation connects the low lying western Grand Valley
area through the lower mountains north of the San Juan chain to the
San Luis Valley in southern Colorado.
A~similar pattern is present for the cross-correlation function,
which is distinct from the summer behavior which indicated
near-independence between minimum and maximum temperatures over the complex
topography.

%
\begin{figure}

\includegraphics{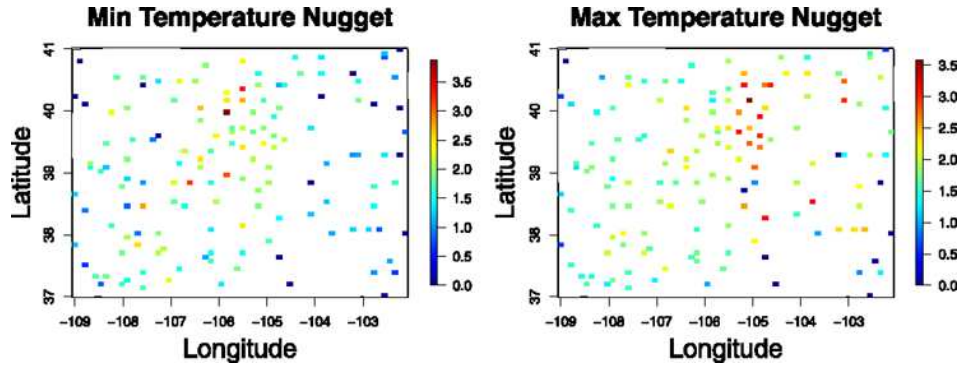}

\caption{Estimated nuggets $\hat{\tau}_N(\bs)$ and $\hat{\tau
}_X(\bs)$
at observation network locations, units are degrees Celsius.} \label
{fignuggets}
\end{figure}


A notable departure of our model from typical geostatistical approaches
is in allowing the nugget effect to vary with location. Our motivation
is that the small scale spatial structure is expected to be dampened
in the eastern plains with stable orography, but potentially inflated
over the mountainous region of Colorado. Figure~\ref{fignuggets} displays
the local estimates $\hat{\tau}_i(\bs)$ for $i=N,X$ at the locations within
our observation network. For both minimum and maximum temperatures, the
nugget effects tend to be less over the eastern plains, indicating
less fine scale spatial structure (although there is yet some evidence
of small scale structure in the maximum temperature nuggets here).
Over the Rocky Mountains, especially the northern Rockies, minimum
temperature exhibits inflated nugget effects, indicating fine scale
spatial processes in the complex terrain. Similarly, the finest scale
spatial structures indicated by these nugget effects for maximum
temperature fall almost directly along the front range, the longitude
line of approximately $-105^\circ$, indicating highly variable
maximum temperatures between the boundary of the plains and sudden
mountainous terrain. The inclusion of a spatially varying $\tau_i(\bs)$
allows the statistical model to retain increased variability along
the front range, for example, while simultaneously generating tempered
fields over the eastern plains and fields of medium variability over
the main Rockies and western slopes.

%
\begin{figure}

\includegraphics{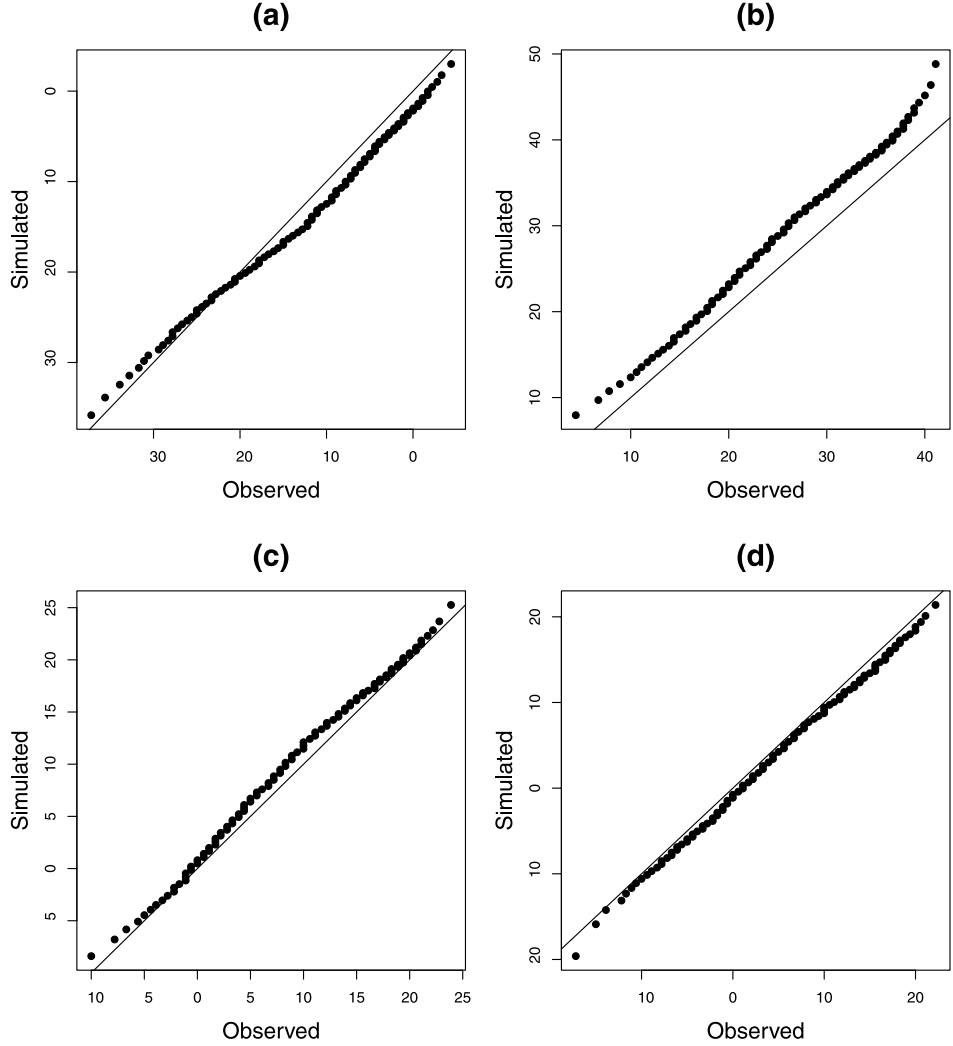}

\caption{Q--Q plots for daily spatial extrema, comparing \textup{(a)}
domain minimum
of minimum temperature, \textup{(b)} domain maximum of maximum temperature,
\textup{(c)} domain maximum of minimum temperature and \textup{(d)}
domain minimum
of maximum temperature, units are degrees Celsius.}
\label{figextrema}\vspace*{6pt}
\end{figure}
%

An increasingly important consideration in climate science is the
effect of climate change on extremes [\citet{easterling2000}]. Our
model is not
explicitly designed to replicate extreme events, as we focus
mainly on the first and second order properties of minimum and maximum
temperatures. Figure~\ref{figextrema} shows Q--Q plots for daily domain-wide
extrema. In particular, we find the minimal and maximal
domain-wide temperatures
$Z_{i,\min}(t)=\min_{\mathbf{s}}\{Z_i(\bs,t)\}$ and
$Z_{i,\max}(t)=\max_{\mathbf{s}}\{Z_i(\bs,t)\}$, and compare
simulated to observed daily statistics for $i=N,X$. Our model
replicates the statistical properties of $Z_{N,\min}(t), Z_{X,\min}(t)$
and $Z_{N,\max}(t)$ very well, at even the most extreme tails of
these domain extrema. However, we simulate domain-wide maximal
maximum temperatures that are slightly too high, on average
about $2^\circ$C.
Overall, even though our approach does not explicitly model extreme
temperatures, we are able to capture the spatial extrema with
reasonable accuracy.

%
\begin{figure}

\includegraphics{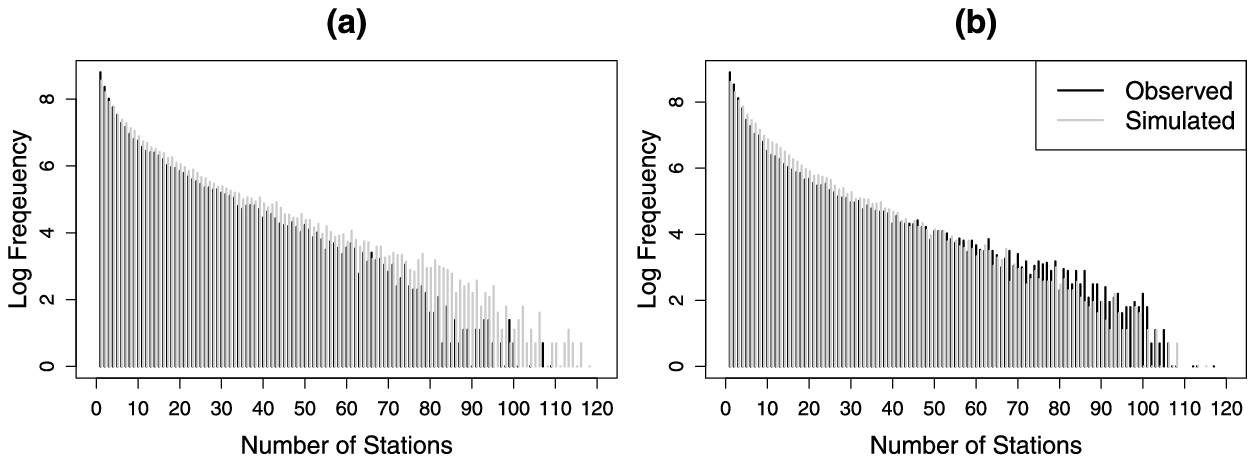}

\caption{Log frequency of observed and simulated local residual
threshold exceedances. Each bar's height is the log freqeuency
(i.e.,~log number of days) that
an exact number of the observation network stations had
weather that exceeded the local $90\%$ quantile for \textup{(a)}
maximum temperature,
or whose weather fell below the local $10\%$ quantile
for \textup{(b)} minimum temperature.}\label{figexceedance}
\end{figure}


While our model adequately replicates domain-wide extrema, the
related quantity of spatially consistent local extrema is critically
important to replicate. In particular, for energy use forecasting
and modeling, if a large number of locations experience unusually
low or high temperatures simultaneously, then the load on the
energy grid can be much greater than if the temperature anomaly
were highly localized. Figure~\ref{figexceedance} shows
log frequencies (i.e.,~the log number of days) of the number of stations
whose local weather
process $W_i(\bs,t)$ either exceeded the local $90\%$ quantile
(i.e., the quantile using only data from location $\bs$)
or fell below the local $10\%$ quantile, corresponding to local
hot or cold events, respectively. Our approach captures
the spatial frequencies of unusual local cold temperatures extremely
well, and tends to simulate local heat events over slightly
inflated regions when many stations experience hot events, although
usually fewer than seven extra days on average.

%
\begin{figure}

\includegraphics{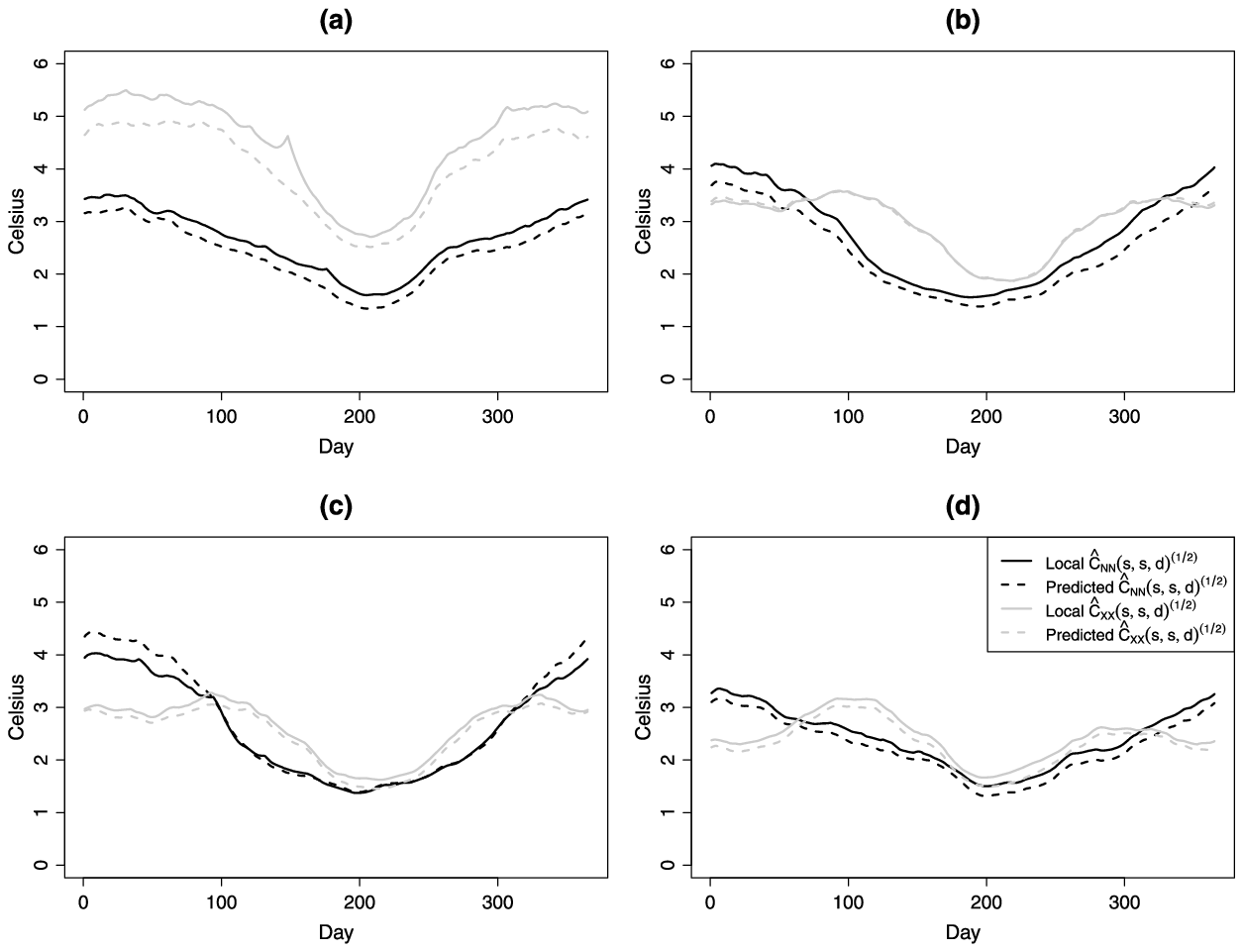}

\caption{Locally estimated standard deviations
$(\hat{C}_{ii}(\bs,\bs,d)^{1/2})$
for $i=N,X$ on all days of the calendar year $d=1,\ldots,365$,
and predicted standard deviations for the four hold out stations
$\bs= $
\textup{(a)} Kit Carson, \textup{(b)} Estes Park, \textup{(c)} Buena
Vista and \textup{(d)} Delta.}
\label{figSDCV}
\end{figure}


Our nonparametric weather component covariance estimator (\ref{eqnp})
is not optimized for cross-validation. To assess the interpolative
properties of our estimator, we hold out data from the four network
stations shown in Figure~\ref{fignetwork}, representing four
distinct regimes of Colorado.
We predict the local standard deviations $\hat{C}_{ii}(\bs,\bs,d)^{1/2}$
for $i=N,X$ and compare these to the locally estimated
values of $\hat{C}_{ii}(\bs,\bs,d)^{1/2}$ when station data
is retained. Figure~\ref{figSDCV} contains the local and predicted
estimates for all days of the calendar year. Clearly the weather
component variability is highly dependent on season as well as
location, particularly for maximum temperature there is substantially
greater variability in the eastern plains ($3^\circ$--$5^\circ$C) compared
to the mountain regions ($2^\circ$--$3.5^\circ$C).
Our predictive local standard deviations (dashed lines in Figure~\ref{figSDCV})
generally agree closely with the local estimates, although there is
a slight tendency to under-predict local standard deviation at
Kit Carson by $0.1^\circ$--$0.3^\circ$C.
Not only are the raw values well predicted,
but the climatological curvature is preserved as well; for example,
we successfully replicate the increased variability of maximum
temperature over the western slopes during springtime with relatively
constant variability throughout the three remaining seasons (panel d)
while simultaneously producing significant seasonality over the
eastern plains, with low variability during summer and high variability
during winter (panel a).

%
\begin{table}
\tabcolsep=0pt
\caption{Interpolated estimates (with predictive standard deviation)
of the local
climate component coefficients with the validating locally estimated parameters.
Locations are $\bs= $ \textup{(a)}~Kit Carson, \textup{(b)}~Estes
Park, \textup{(c)}~Buena Vista and \textup{(d)}~Delta.
Predictions are starred if the truth is outside of the predictive $95\%
$ confidence
interval. Units are degrees Celsius for $\beta_0, \beta_1$ and $\beta
_2$, unitless for
$\beta_3$ and $\beta_4$, and degrees Celsius per century for $\beta
_5$}\label{tabests}
\begin{tabular*}{\textwidth}{@{\extracolsep{\fill
}}ld{2.9}d{2.8}d{2.9}d{2.8}d{2.2}d{2.2}d{2.2}d{2.2}@{}}
\hline
& \multicolumn{4}{c}{\textbf{Kriged estimate (kriging standard
deviation)}} & \multicolumn{4}{c@{}}{\textbf{Local estimate}}
\\[-6pt]
& \multicolumn{4}{c}{\hrulefill} & \multicolumn{4}{c@{}}{\hrulefill
} \\
& \multicolumn{1}{c}{\textbf{a}} & \multicolumn{1}{c}{\textbf{b}} &
\multicolumn{1}{c}{\textbf{c}} & \multicolumn{1}{c}{\textbf{d}} &
\multicolumn{1}{c}{\textbf{a}} & \multicolumn{1}{c}{\textbf{b}} &
\multicolumn{1}{c}{\textbf{c}} & \multicolumn{1}{c@{}}{\textbf{d}}
\\
\hline
$\beta_{0N}(\bs)$ & -2.70\ (1.40) & -5.43\ (1.32) & -6.37\ (1.35) &
-4.39\ (1.34) &
-3.43 & -4.76 & -4.58 & -4.49 \\
$\beta_{1N}(\bs)$ & -3.94\ (0.45) & -2.45\ (0.41) & -2.89\ (0.43) &
-1.78\ (0.42) &
-4.48 & -2.01 & -2.66 & -1.69 \\
$\beta_{2N}(\bs)$ & -1.08\ (0.25) & -0.63\ (0.23) & -0.81\ (0.24) &
-0.36\ (0.24) &
-1.07 & -0.68 & -0.58 & -0.19 \\
$\beta_{3N}(\bs)$ & 0.20\ (0.06) & 0.28\ (0.06) & 0.27\ (0.06) &
0.27\ (0.06) &
0.20 & 0.27 & 0.21 & 0.25 \\
$\beta_{4N}(\bs)$ & 0.46\ (0.06) & 0.41\ (0.06) & 0.47\ (0.06) &
0.50\ (0.06) &
0.45 & 0.38 & 0.49 & 0.52 \\
$\beta_{5N}(\bs)$ & 0.55\ (1.30) & 0.53\ (1.30) & 0.57\ (1.30) &
0.60\ (1.30) &
0.42 & 0.67 & 0.58 & 0.72 \\[3pt]
$\beta_{0X}(\bs)$ & 6.99\ (1.08) & 4.40\ (0.98) & 3.27\ (1.02) &
4.31\ (1.01) &
7.28 & 4.03 & 3.62 & 4.36 \\
$\beta_{1X}(\bs)$ & -4.30^*\ (0.41) & -3.80\ (0.38) & -3.81^*\ (0.39)
& -3.77\ (0.39) &
-5.29 & -3.30 & -3.00 & -4.08 \\
$\beta_{2X}(\bs)$ & -1.26^*\ (0.15) & -1.31\ (0.14) & -1.29^*\ (0.14)
& -0.93\ (0.14) &
-1.58 & -1.20 & -0.93 & -0.90 \\
$\beta_{3X}(\bs)$ & 0.03\ (0.07) & -0.01\ (0.07) & -0.02\ (0.07) &
-0.03\ (0.07) &
-0.05 & -0.01 & -0.02 & -0.08 \\
$\beta_{4X}(\bs)$ & 0.63\ (0.05) & 0.67\ (0.05) & 0.69\ (0.05) &
0.75\ (0.05) &
0.64 & 0.70 & 0.75 & 0.78 \\
$\beta_{5X}(\bs)$ & 0.30\ (0.80) & 0.31\ (0.80) & 0.31\ (0.80) &
0.30\ (0.80) &
-0.42 & 0.67 & 0.48 & 0.13 \\
\hline
\end{tabular*}
\end{table}
%

Table~\ref{tabests} shows the interpolated coefficients with predictive
standard deviation, along with the locally estimated parameters $\hat
{\beta}_{ki}(\bs)$
for $k=0,\ldots,5$ and $i=N,X$ for $\bs$ being one of the four held out
network stations. All locally estimated parameters are within the
$95\%$ predictive confidence interval, except for four cases for maximum
temperature. Our predictive intervals are calibrated; the coverage of
the $95\%$ interpolation
intervals for leave-one-station-out cross-validation over all locations
was, at worst, $92.4\%$ for $\beta_{2X}(\bs)$.
Notice that the local estimates vary substantially between
locations, indicating that indeed the local climate varies over the domain.
Hence, we are able to successfully predict the local weather component
parameters
and local climate component parameters at these four hold out locations which
are representative of four regimes in Colorado.


%
\begin{figure}

\includegraphics{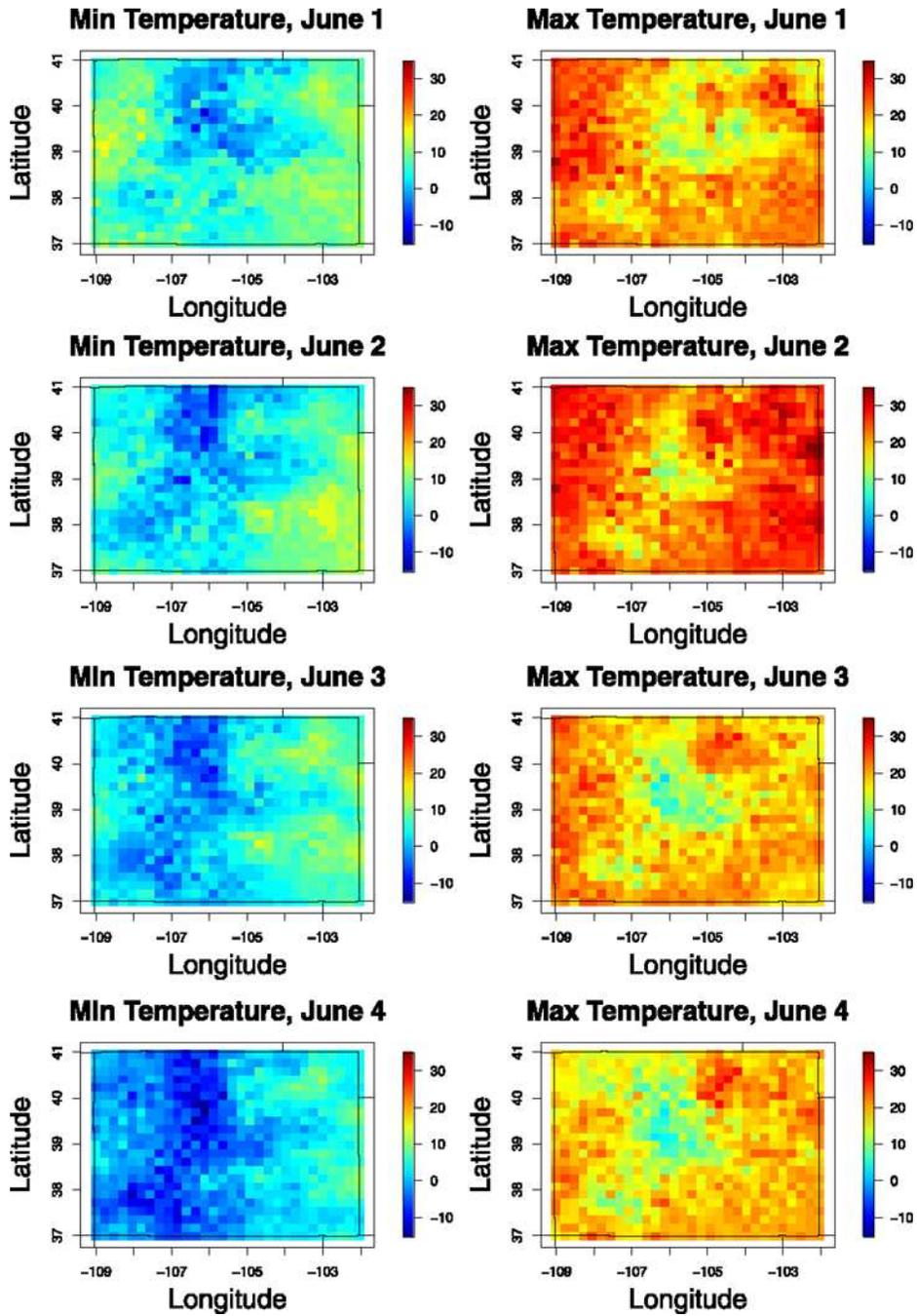}

\caption{Gridded simulation of daily minimum and maximum
temperatures on days June~1--4.} \label{figsim}
\end{figure}


Finally, we illustrate the final product of our approach in Figure
\ref{figsim} which displays four days of gridded simulations of
minimum and maximum temperatures over Colorado. Marginally, we
visually see the temporal persistence of temperature over a period
of days, as both minimum and maximum temperatures experience a period
of cooling over June 1--4. Notice the effect of local climate is
to keep the Rocky Mountain region cooler for both variables, allowing
higher minimum and maximum temperatures to fall over the eastern plains
of Colorado. We also see slightly warmer temperatures on the western
slopes, as the Rocky Mountains decay in elevation to the western
border of Colorado. The cross-correlation between the two variables
is also present, as both variables are seen to cool across the domain
simultaneously.

\section{Discussion} \label{secdiscussion}

In this paper we introduce a framework for stochastic bivariate minimum
and maximum
temperature simulation over complex domains. The framework
distinguishes between
local climate and weather processes. The local climate is accommodated
through a linear model whose coefficients are spatially varying, and
the weather process is modeled as a bivariate spatial Gaussian process with
a nonparametric estimate of the matrix-valued covariance function that
retains nonnegative definiteness at arbitrary locations. We successfully
capture the temporally varying spatial dependence between minimum and
maximum temperatures over the state of Colorado, which exhibits challenging
complex terrain that is difficult for extant models to accommodate.

Our nonparametric estimator smooths multivariate spatial covariance
over space
as well as time. This approach allows spatial dependence to be highly different
during winter than during summer, for instance, and also retains nonstationary
spatial structures both within each process and between processes. The
estimator is available at any location, not only those within the observation
network, and always retains nonnegative definiteness, allowing for gridded
simulations. The estimator relies on kernel-smoothed empirical covariance
functions, and our current approach to spatial bandwidth selection is
ad hoc.
One future route of research may be to decide on a quantitative
approach to
bandwidth selection when sharp boundaries and highly variable covariances
are expected across the study domain, notably different than most mean
function smoothing literature [\citet{wand1995}].
A second potential direction of research may be to develop a nonparametric
kernel-smoothed estimate of the multivariate covariance function that is
robust against outliers and still retains nonnegative definiteness.

While our approach does not explicitly model extremes, our simulations
indicate reasonable replication of tail behavior, even domain-wide
extrema. A limitation in using Gaussian processes
is that there is a lack of clustering at high levels, both spatially
and temporally [\citet{sibuya1960}]. This is one potential explanation
for the behavior of Figure~\ref{figextrema}(b), where domain-wide
maximal maximum temperatures were simulated slightly above the
observed extremes, although we would expect to see similar behavior
in panels (a), (c) and (d). An approach that includes a Gaussian
process model for the bulk of the distribution along with a model
for spatial extremes may improve extremal performance.

One consideration of our model is that we do not explicitly force
the simulation of maximum temperature to be greater than or equal to
minimum temperature; in our Colorado example maximum temperature was
less than minimum temperature for approximately one tenth of a percent of
our simulations. It may be of interest to adopt the models of
\citet{jolliffe1996} or \citet{jones2004} to our situation if this
issue is of critical concern.

The clearest route of future research is to extend our ideas to a full
stochastic weather simulator that can simulate spatially correlated
fields of
multiple variables such as minimum and maximum temperatures,
precipitation amount, solar radiation, wind direction/wind speed and relative
humidity simultaneously. Indeed, in complex terrains the practitioner will
need to rely on highly flexible spatial models to replicate the strong
nonstationarities exhibited by these various processes, as well as
the complicated spatially evolving relationship between them.

\begin{appendix}
\section*{Appendix}\label{app}

In this Appendix we show the nonparametric estimator (\ref{eqnpR}) is
nonnegative definite, from which it follows that (\ref{eqnp}) and
(\ref{eqnpmissing})
are also nonnegative definite. Below, we present an argument that (\ref{eqnp})
is asymptotically unbiased for $C_{ij}(\mathbf{x},\by,d(t_0))$.

The nonnegative definiteness property is not restricted to a bivariate
process, so assume there are $p$ spatial processes $W_i(\bs,t),
i=1,\ldots,p$,
with observation network locations $\bs_m, m=1,\ldots,n$.
Define $U_i(\bs,t) = \break W_i(\bs,t)
\indicator_{[W_i(\mathbf{s},t)\ \mathrm{is\ observed}]}$, noting that
if $W_i(\bs,t)$ is unavailable at a particular location and time,
$U_i(\bs,t) = 0$.

Consider evaluating $R_{ij}(\mathbf{x}_k,\mathbf{x}_\ell,t)$ at any arbitrary
locations $\mathbf{x}_k$ and $\mathbf{x}_\ell,\break k,\ell=1,\ldots,N$, and define the
arbitrary vector $\ba= (a_{11},\ldots,a_{1N},a_{21},\ldots,a_{pN}).$
Set $\Sigma$ to be the covariance matrix made up of the functions
$R_{ij}(\cdot,\cdot,t)$ corresponding to the random vector
\[
\bigl(W_1(\mathbf{x}_1,t),W_1(
\mathbf{x}_2,t),\ldots,W_p(\mathbf{x}_N,t)
\bigr)'.
\]
Then, absorbing the denominator into the kernel functions of
$R_{ij}(\mathbf{x}_k,\mathbf{x}_\ell,t)$, and writing $R_{ij}'
(\mathbf{x}_k,\mathbf{x}_\ell,t)$
for this normalized function, we have
\begin{eqnarray*}
\ba' \Sigma\ba&=& \sum_{i,j=1}^p
\sum_{k,\ell=1}^N a_{ik}
a_{j\ell} R_{ji}'(\mathbf{x}_\ell,
\mathbf{x}_k,t)
\\
&= &\sum_{i,j=1}^p \sum
_{k,\ell=1}^N a_{ik} a_{j\ell} \sum
_{m,r=1}^n K_\lambda\bigl(\|
\mathbf{x}_\ell- \bs_m\|\bigr) K_\lambda\bigl(\|\mathbf{x}_k -
\bs_r\|\bigr) U_i(\bs_r,t) U_j(
\bs_m,t)
\\
&= &\sum_{m,r=1}^n \sum
_{i,j=1}^p \sum_{k,\ell=1}^N
\bigl(a_{ik} K_\lambda\bigl(\|\mathbf{x}_k - \bs_r
\|\bigr) U_i(\bs_r,t) \bigr) \bigl(a_{j\ell}
K_\lambda\bigl(\|\mathbf{x}_\ell- \bs_m\|\bigr) U_j(
\bs_m,t) \bigr)
\\
&= &\Biggl( \sum_{r=1}^n \sum
_{i=1}^p \sum_{k=1}^N
a_{ik} K_\lambda\bigl(\|\mathbf{x}_k - \bs_r\|\bigr)
U_i(\bs_r,t) \Biggr)^2 \geq0.
\end{eqnarray*}

To show that (\ref{eqnp}) is asymptotically unbiased for $C_{ij}(\mathbf{x},\by,d(t_0))$,
we disregard the smoothing over time, since asymptotically we do not have
a finer resolution of time points (but for consistency we would assume an
increasing number of realizations per each day of the year).
In particular, suppose we observe the bivariate process $(W_N(\bs
_k),W_X(\bs_k))$
for $\bs_1,\ldots,\bs_n\in\mathcal{D} \subset\real^d$, which
are samples from a distribution with strictly positive probability density
$f\dvtx\mathcal{D}\rightarrow\real^+$, with empirical c.d.f.
$F_n(\mathbf{x}) =
\frac{1}{n}\sum_{k=1}^n \indicator_{[\mathbf{s}_k \leq
\mathbf{x} ]}$,
where the indicator function is 1 if the inequality holds for all indices
of $\mathbf{x}$. The density~$f$, with
corresponding c.d.f. $F$,
allows the network density to vary across the domain.
We additionally suppose $n\rightarrow\infty$ and $\lambda\rightarrow0$
such that $\lambda\sim n^{-1/d + \varepsilon}$ for some small
$0<\varepsilon<1/d^2$.

Suppressing the time indexing from our notation, we can write
\[
\hat{C}_{ij}(\mathbf{x},\by) = \frac
{1}{n^2\lambda^{2d}}\sum
_{k=1}^n \sum_{\ell=1}^n
K_\lambda' \bigl(\|\mathbf{x}-\bs_k\| \bigr) K_\lambda'
\bigl(\|\by-\bs_\ell\| \bigr) W_i(\bs_k)
W_j(\bs_\ell),
\]
where the denominator of (\ref{eqnp}) is absorbed into the kernel
functions of the numerator, yielding standardized functions $K_\lambda'$.
Here we only consider the direct covariance estimators at a location
$\mathbf{x}\in{\cal D}\setminus\partial{\cal
D}$, $C_{ii}(\mathbf{x},\mathbf{x})$; the same
argument applies for the direct and cross-covariance functions
$C_{ij}(\mathbf{x},\by)$
for $\mathbf{x}\not=\by$. We have
%
\begin{eqnarray}
\E\hat{C}_{ii}(\mathbf{x},\mathbf{x}) &=& \frac{1}{n^2 \lambda^{2d}} \sum
_{k,\ell=1}^n K_\lambda'\bigl (\|\mathbf{x}-
\bs_k\| \bigr)K_\lambda' \bigl(\|\mathbf{x}-\bs_\ell\| \bigr)
\E\bigl(W_i(\bs_k) W_i(\bs_\ell)
\bigr)
\\
&=& \frac{1}{n^2 \lambda^{2d}} \sum_{k,\ell=1}^n
K_\lambda' \bigl(\|\mathbf{x}-\bs_k\| \bigr)K_\lambda'
\bigl(\|\mathbf{x}-\bs_\ell\| \bigr) C_{ii}(\bs_k,
\bs_\ell)
\\
&&{} + \frac{1}{n^2 \lambda^{2d}} \sum_{k = 1}^n
K_\lambda' \bigl(\|\mathbf{x}-\bs_k\| \bigr)^2
\tau_i(\bs_k,\bs_k)^2.
\end{eqnarray}
Invoking Lemma 7 of \citet{kleiber2012nsmm}, in the limit as
$n\rightarrow\infty$,
we can pass from the sum to the integral. Assume the empirical c.d.f.
$F_n$ is
close to the limiting c.d.f. $F$, where $\sup_{\mathbf{x}}
|F_n(\mathbf{x}) - F(\mathbf{x})| =
D_n$ where $D_n = o(1/(n\lambda^d))$. This rate holds, for example,
if ${\cal D} = [0,1]$, $F$ is the uniform density and $F_n$ is the empirical
c.d.f. of the uniform grid $(1/n,2/n,\ldots,n/n)$. For $d > 1$, if $n$
grows as $M^d$, a rate of $D_n \sim1/n^{1/d}$ can be derived for
sampling locations on a regular grid with limiting uniform distribution
[\citet{kleiber2012nsmm}]. Then we have
%
\begin{eqnarray}
\E\hat{C}_{ii}(\mathbf{x},\mathbf{x}) &= &\frac{1}{\lambda^{2d}} \iint_{{\cal D}^2}
K_\lambda' \bigl(\|\bu-\mathbf{x}\| \bigr)K_\lambda' \bigl(\|
\bv-\mathbf{x}\| \bigr) C_{ii}(\bu,\bv) \,\mathrm{ d}F(\bu
)\,\mathrm{ d}F(\bv)
\nonumber
\\[-6pt]
\\[-10pt]
\nonumber
& &{}+ \frac{1}{n \lambda^{2d}} \int_{{\cal D}} K_\lambda'
\bigl(\|\bu-\mathbf{x}\| \bigr)^2 \tau_i(\bu,\bu)^2 \,\mathrm{ d}F(
\bu) + {\cal O}(D_n).
\end{eqnarray}
Making the change of variables to $\ba= (\bu- \mathbf{x})/\lambda$ and
$\bb= (\bv- \mathbf{x})/\lambda$ yields
%
\begin{eqnarray}
\label{eqasymptoticbias} &&\iint_{{\cal D'}^2} K'\bigl (\|\ba\|
\bigr)K'\bigl (\|\bb\| \bigr) C_{ii}(\lambda\ba+ \mathbf{x},\lambda\bb+ \mathbf{x}) {
\,\mathrm d}F(\ba) \,\mathrm{ d}F(\bb)
\nonumber
\\[-8pt]
\\[-8pt]
\nonumber
&&\qquad{} + \frac{1}{n \lambda^{d}} \int_{{\cal D'}} K' \bigl(\|\ba\|
\bigr)^2 \tau_{i}(\lambda\ba+ \mathbf{x},\lambda\ba+
\mathbf{x})^2\,\mathrm{ d}F(\ba) + {\cal O}(D_n)
\end{eqnarray}
for an appropriate translated domain ${\cal D'}$.
As $\lambda\sim n^{-1/d + \varepsilon}$,
the second term of (\ref{eqasymptoticbias}) converges to zero.
The arguments from \citet{kleiber2012nsmm} applied to the first term of
(\ref{eqasymptoticbias}) then yield the unbiasedness of $C_{ii}(\mathbf{x},\mathbf{x})$.
\end{appendix}


%


\printaddresses

\end{document}